\documentclass[%
 reprint,
 amsmath,amssymb,
 aps,
]{revtex4-1}

\usepackage{graphicx}
\usepackage{dcolumn}
\usepackage{bm}
\usepackage[usenames, dvipsnames]{color}

\begin{document}

\preprint{APS/123-QED}

\title{Anisotropic Mimetic Cosmology}

\author{M. H. Abbassi}%
 \email{mh.abbassi@gmail.com}
\affiliation{%
Department of Physics, School of Science, Tarbiat Modares University. P.O. Box 14155-4838, Tehran, Iran
}%

\author{A. Jozani}
 \email{jozani.alireza@gmail.com}
\affiliation{
Department of Physics, Shahid Beheshti University, G.C., Evin, Tehran 19839, Iran
}%

\author{H. R. Sepangi}
 \email{hr-sepangi@sbu.ac.ir}
\affiliation{%
Department of Physics, Shahid Beheshti University, G.C., Evin, Tehran 19839, Iran
}%

\date{\today}

\begin{abstract}
We consider a mimetic set up in which the mimetic scalar is coupled to a vector field. It is shown that such a field with a time-like component does not contribute to the background equations and yet produces healthy isocurvature perturbations with respect to ghost and gradient instabilities in spite of the absence of any propagating curvature perturbations at the level of the quadratic action. We then consider a vector field with space-like components which leads to an anisotropic Bianchi universe and show that the ghost and gradient instabilities are absent in the limit of high momenta and that the propagating curvature perturbations have healthy UV behavior.
\end{abstract}

\maketitle


\section{\label{intro} Introduction}

The dawn of mimetic theory has its roots in \cite{Chamseddine:2013kea} where it was shown that a carefully designed  conformal transformation of the metric would mimic the behavior of Dark Matter (DM). That is, if one writes the physical metric as $g_{ab}^{phys}=X \bar{g}_{ab}$, where $X=g^{ab}\partial{_a}\phi\partial{_b}\phi$ and $\bar{g}_{ab}$ is an auxiliary metric, then taking variation with respect to the auxiliary metric results in the equation of motions (EOM) only in terms of the physical metric and scalar field. The extra terms in the EOM behave exactly as the matter content. In this way it was shown that such a theory of gravity with a special kind of conformal degrees of freedom can mimic the DM behavior. This form of conformal transformation imposes a constraint on the scalar field, namely $g^{ab}\partial_a\phi\partial_b\phi=-1$. However, it was later shown that \cite{Golovnev:2013jxa} this result is equivalent to a theory where such a constraint is imposed via a Lagrange multiplier in the action. Therefore, the action for mimetic gravity can simply be written as
\begin{equation}
S=\int d^4x \sqrt{-g}\left(\frac{1}{2} \mathcal{R}+\lambda(t)(g^{ab}\partial_a\phi\partial_b\phi+1)\right).
\end{equation}
For a FRW background
\begin{equation}
ds^2=-dt^2+a(t)^2 d\vec{x}^2,
\end{equation}
the mimetic constraint causes the scalar field in a homogeneous background to obey a simple solution of the form $\phi=t+\mbox{const}$. The metric and scalar field equations of motion then become
\begin{eqnarray}\label{mimeom}
3 H(t)^2=&-2\lambda(t),\\ \label{eq:tt}
2\dot{H}(t)+3H(t)^2=&0,\\ \label{eq:ii}
2\lambda(t)=&-\frac{c}{a(t)^3}.
\end{eqnarray}
It is clear from these background equations that the scalar degree of freedom behaves exactly as Dark Matter. In spite of such behavior in the background, it was shown that although it is free from instabilities, it does not have propagating degrees of freedom \citep{Barvinsky:2013mea}. This means that the speed of sound for perturbations is zero and therefore this new proposed degree of freedom does not contribute to the formation of structures in the universe. To remedy such a flaw, the authors of \cite{Chamseddine:2014vna,Capela:2014xta,Mirzagholi:2014ifa,Chamseddine:2016uef} added a term of the form $\gamma (\Box\phi)^2$ to the above action. This new term does not change the DM-like behavior of the scalar degrees of freedom and produces a constant nonzero speed of sound. Unfortunately, as it turned out, this new action also suffers from  ghost or gradient instabilities \citep{Ijjas:2016pad,Ramazanov:2016xhp,Firouzjahi:2017txv}. Subsequently, a plethora of research work have appeared whose goal is to construct a mimetic DM theory without such shortcomings \cite{Hirano:2017zox,Zheng:2017qfs, Gorji:2017cai,Takahashi:2017pje}, mostly to no avail in that non of which has been completely successful in getting an exact DM behavior through a healthy stable theory. For more extensive discussions about other aspects of mimetic gravity look at \cite{Sebastiani:2016ras}.

In addition to Dark Matter, our universe consists of another unknown component called Dark Energy (DE), necessary to explain the present accelerating expansion of the observed universe. There are several proposals to explain DE, ranging from the cosmological constant to modifying Einstein theory of gravity. Among different  approaches to describe DE, inclusion of vector degrees of freedom \cite{ArmendarizPicon:2004pm,Wei:2006tn,Jimenez:2008au,Haghani:2016oxv} has proved beneficial, motivated in part by the inflationary paradigm \cite{Gumrukcuoglu:2007bx,Koivisto:2008xf,Dulaney:2010,Gumrukcuoglu:2010yc,Emami:2013bk}. Although there are some cases where one may engineer a model such as to avoid the anisotropy caused by the  vector field in the background,  introduction of a vector field generally causes an anisotropy in the universe which is highly constrained by observational data. However, aside from an observational point of view, it is interesting to investigate the cosmological properties of an anisotropic universe.

As was mentioned above, for consistency of a model,  stability is  one of the most important criteria which should be satisfied. To this end, one imposes  small perturbations over the background, $\phi^{(I)}=\bar{\phi}^{(I)}+\delta\phi^{(I)}$, where index $(I)$ represents different fields in the problem and looks for any pathology in its behavior  \cite{Rubakov:2014jja}. There are two possible methods to investigate this issue. One is to use first order equations of motion of perturbations and to find the solutions to see if they  are not growing. Although this method offers a good sense on gradient instability, it would be misleading for ghost instability, as was shown to be the case in \cite{Chamseddine:2014vna}. The other  method is to write the action up to second order in perturbations. Taking into account that linearized EOMs mostly have second order derivatives of the modes, a generic Lagrangian for such a system can be written as
\begin{equation}
\begin{split}
S=\int &d^4x\biggl(\frac{1}{2}\bold{K}_{(IJ)}\delta\dot{\phi}^{(I)}\delta\dot{\phi}^{(J)} -\bold{G}_{(IJ)}\partial_i\delta\phi^{(I)}\partial^i\delta\phi^{(J)}\\
&-\frac{1}{2}\bold{V}_{(IJ)}\delta\phi^{(I)}\delta\phi^{(J)}\biggr).
\end{split}
\end{equation}
A solution would be stable if
\begin{equation}\label{stab}
\bold{K}>0, \hspace{2mm}\bold{G}>0,\hspace{2mm} \bold{V}>0,
\end{equation}
which means that the matrices $\bold{K}$, $\bold{G}$ and $\bold{V}$ should be positive definite. The first inequality above is the condition to avoid ghosts, the second corresponds to gradient instability and the last is to make the theory free from tachyons. Diagonalizing these matrices and factoring  $\bold{K}$ out, one defines $c_s^2=\frac{\bold{G}}{\bold{K}}$, which is the well known speed of sound.

In this work we investigate the mimetic theory by adding vector degrees of freedom to the action which directly couples to the mimetic field. The goal is to see if this new degree of freedom leads to healthy propagating perturbations and still retaining DM behavior.

The organization of the paper is as follows: In section II we introduce the form of our Lagrangian and derive the equations of motion. In section III we analyze the case for a time-like vector field which corresponds to a FRW background. Next  we study the case of a space-like vector field in section IV which leads to an anisotropic Bianchi background. In section V perturbations are performed on a Bianchi background to check their stability and finally, conclusions are drawn in the last section. Through out the paper we use the signature $(-+++)$.
\section{Mimetic Scalar-Vector Model}

To start with, we consider the following action
\begin{eqnarray}\label{action}
\begin{split}
S=\int d^4x \sqrt{-g} &\biggl(\frac{1}{2}\mathcal{R}+\gamma F_{ab}F^{ab}+\eta\Box\phi A_aA^a\\
&+\lambda(t)(\partial_a\phi\partial^a\phi+1)\biggr).
\end{split}
\end{eqnarray}
Here, the second term is the kinetic term for the vector field $A_a$ and we have imposed the usual mimetic constraint on the scalar field $\phi$ with an interaction represented by the third term. This is a Proca-like term for the vector field where the mass is proportional $\Box{\phi}$. This term breaks the $U(1)$ symmetry of the vector field in the action.

It is also worth mentioning that introduction of a $\Box \phi$ term does not lead a Ostrogradsky kind instability because the mimetic constraint forces higher derivatives of the degree of freedom to vanish, e.g. terms like $\ddot{\phi}$ should be absent. In principle, one could add other operators to the action, but such terms only  contain $\phi$ and since $\phi=t$ in a mimetic setting they are not dynamically interesting. Other operators could also be considered containing a first derivative of the field, like $\partial_a\phi\partial_b\phi A^aA^b$.  However, a dimensional analysis reveals that they are suppressed relative to the other terms in the action.

Variation of the action with respect to $g_{ab}$ gives
\begin{equation}\label{EomFRWg}
\begin{split}
&\mathcal{R}_{ab}-\frac{1}{2}g_{ab}\mathcal{R}=-4\gamma (F_{ac}F^{c}_{b}- \frac{1}{4}g_{ab}F_{cd}F^{cd})\\
&+2\eta A^c(\nabla_a A_c \partial_{b}\phi+\nabla_b A_c \partial_{a}\phi-g_{ab}\nabla^d A_c \partial_d \phi)\\
&-2\eta \Box{\phi}A_{a}A_{b}-2\lambda\partial_a\phi\partial_b\phi,
\end{split}
\end{equation}
while varying the action with respect to the scalar and vector fields $\phi, A_a$ leads to
\begin{equation}\label{EomFRWphi}
\lambda \Box{\phi}+\partial_a\phi\partial^a\lambda=\eta(A_a\Box A^a+\nabla_a A_b\nabla^a A^b),
\end{equation}
and
\begin{equation}\label{EomFRWA}
\nabla_b F^b_a=-\frac{1}{2}A_{a}\Box{\phi},
\end{equation}
respectively. Finally, variation with respect to $\lambda$ results in the usual mimetic constraint, $\partial_a\phi\partial^{a}\phi=-1$. This simply forces the scalar field to be of the form $\phi=t$, where we have ignored the constant by a time re-parametrization. We note that equation (\ref{EomFRWphi}) is a total derivative
\begin{equation}
\nabla_a(\lambda \partial^a\phi-\eta A_b \nabla_a A^b)=0,
\end{equation}
which would allow one to compute $\lambda$ analytically. As we shall see in the following sections, $\lambda$ is represented by two terms, one is of a matter type density $\frac{c}{a^3}$ and the other results from the vector field. This particularly motivates us to take the interaction term as $\Box{\phi}$ coupled to the mass term of the vector field.

\section{Time-like Vector Field:\\FRW Universe }

To get an isotropic universe in the presence of a vector field in the background we may consider $\langle A_a\rangle=(0,\vec{0})$ or a time-like vector field represented by $A_a=(A_0(t),\vec{0})$. If the second is considered then the zero component of equation (\ref{EomFRWA}) results in $3\eta H A_0=0$, so the latter case would be equal to the former. It is then  clear that the solutions of equations (\ref{EomFRWg}) and (\ref{EomFRWphi}) are similar to equation (\ref{mimeom}). This means that equations of the background reduce to the form of simple mimetic gravity.

Although we arrived at a trivial background (the same as that of the original mimetic theory without any background vector field), it is interesting to see what happens when perturbations are considered. To this and other ends, we write our metric in the ADM form \cite{Maldacena:2002vr}
\begin{equation}
ds^2=-N^2 dt^2+h_{ij}(N^i dt +dx^i)(N^j dt +dx^j),
\end{equation}
where $i,j$ run over spatial dimensions and $N$ and $N_i$ are the usual lapse and shift functions respectively and $h_{ij}$ is the spatial part of the metric. At the level of background we note that $N=1$, $N^i=0$ and $h_{ij}=a(t)^2 \delta_{ij}$. Among the different choices  for the gauge, it is more convenient to do calculations in the co-moving gauge. The special feature of this gauge is that all perturbation are restored in the metric and  perturbation of the scalar field $\phi=\phi(t)+\delta\phi(t,x)$ would be zero, i.e. $\delta\phi(t,x)=0$. It is then possible to parameterize perturbations as follows
\begin{eqnarray}
N=&1+\mathcal{N}(t,\vec{x}),\\
N_i=&-\partial_i \chi(t,\vec{x}) - v_i(t,\vec{x}),\\
h_{ij}=&a(t)^2 e^{2 \zeta(t,\vec{x})} \delta_{ij}. \label{hij}
\end{eqnarray}
Here, $\mathcal{N}$, $\chi$ and $\zeta$ are scalar perturbations and $v_i$ is the vector perturbation which satisfies $\partial_i v_i=0$. It is worth mentioning that we have ignored tensor perturbations in equation (\ref{hij}) in view of the fact that we have only extra scalar and vector degrees of freedom in the Einstein-Hilbert action and therefore their contribution would be trivial. In addition, we have vector field perturbations
\begin{equation}
A_a(t,\vec{x})=\left(\delta A_0(t,\vec{x}),\partial_i\delta A(t,\vec{x})+\delta A_i(t,\vec{x})\right),
\end{equation}
where $\delta A_0$ and $\delta A$ are scalar perturbations and $\delta A_i$ represent vector perturbations which obey $\partial_i A_i=0$. Also, there is the scalar perturbation $\delta\lambda(t,\vec{x})$ for the Lagrange multiplier.

Using Scalar-Vector-Tensor decomposition, we may separately study the scalar and vector perturbations and specifically look  at the second order perturbation of the mimetic constraint
\begin{equation}
\mathcal{L}^{(s)}\supset a^3\mathcal{N}(2\delta\lambda-(\mathcal{N}-6\zeta)\lambda).
\end{equation}
This is the only part in which $\delta \lambda$ appears. The complete form of the quadratic action in real space is given in the appendix. Varying the action with respect to $\delta \lambda$ gives
\begin{equation}
\mathcal{N}(t,\vec{x})=0.
\end{equation}
This is one of the advantages of using the gauge mentioned above in the context of mimetic gravity. It is more convenient to do the rest of calculations in Fourier space. We perform Fourier transformation on each mode according to
\begin{equation}
\delta(t,\vec{x}) = \frac{1}{(2\pi)^3}\int d^3\vec{k}e^{-\vec{k}.\vec{x}}\delta(t,\vec{k}),
\end{equation}
where $\delta$ represents each of the perturbation parameters. We denote the real space and Fourier space values of perturbations with the same symbol $\vec{k}$ which denotes the comoving wave number vector. The physical wave number vector can be defined as
\begin{equation}
\vec{q}\equiv \frac{\vec{k}}{a},
\end{equation}
and we denote its magnitude by $q$. Using $SO(3)$ rotational symmetry, we can go to a frame where $\vec{k}=(0,0,k_z)$, that is, only one longitudinal mode is excited. Taking into account these considerations, the quadratic form of the action  in Fourier Space is given by
\begin{equation}\label{L2}
\begin{split}
\mathcal{L}^{(s)}=& a^3\Bigl(-3\dot{\zeta}^2+q^2\zeta^2+2q^2\chi\dot{\zeta}+(3\eta H-2\gamma q^2)\delta A_0^2\\
&-2\gamma q^2\delta\dot{A}^2-3\eta q^2 H\delta A^2+4q^2 \gamma\delta A_0\delta\dot{A}\Bigr),
\end{split}
\end{equation}
where we have used background equations and integration by parts for further simplification. It is seen that $\chi$ and $\delta A_0$ are two non dynamical degrees of freedom which should be integrated out from the action. Varying (\ref{L2}) with respect to $\delta A_0$ we find
\begin{equation}
\delta A_0=\frac{2\gamma q^2}{2\gamma q^2-3\eta H}\delta\dot{A}.
\end{equation}
Also, variation of (\ref{L2}) with respect to $\chi$ results in $\zeta=\mbox{const}$. As in the original mimetic gravity theory, the curvature perturbation is not propagating. After integrating out the non-dynamical degrees of freedom, our quadratic action of the scalar modes becomes
\begin{equation}\label{L3}
\mathcal{L}^{(s)}= a^3\left(\frac{6 q^2\gamma\eta H}{2\gamma q^2-3\eta H}\delta\dot{A}^2+3q^2\frac{3\eta^2H^2-2q^2\gamma\eta H}{2\gamma q^2-3\eta H}\delta A^2\right).
\end{equation}
Fortunately, this case is tractable analytically in all regimes of momenta.  Comparing coefficients in equation (\ref{L3}) with conditions given in (\ref{stab}), it becomes clear that for $\gamma<0$ and $\eta>0$ the model is healthy with respect to ghost and gradient instabilities.

Let us now concentrate on the vector part of perturbations. As was mentioned earlier, we exploit the $SO(3)$ symmetry of the problem to adapt a reference frame where $\vec{k}=(0,0,k_z)$. In this frame the transverse conditions of the vector modes $\partial_i\delta A_i=0$ and $\partial_i v_i=0$, lead to $\delta A_z=0$, $v_z=0$ and perturbations are given by
\begin{equation}
g_{ab}=
\begin{bmatrix}
-1         & -v_x    &-v_y               & 0\\
-v_x       &         &                   &  \\
-v_y       &         & a^2 \delta_{ij}   &  \\
 0         &         &                   &
\end{bmatrix},
\vec{\delta A}=(0,\delta A_x,\delta A_y,0).
\end{equation}
One may substitute these perturbation in (\ref{action}) and get the quadratic action for vector perturbations
\begin{equation}
\mathcal{L}^{(v)}=a^3 \left(-2\gamma\frac{\delta\dot{\vec{A}}^2}{a^2}+(2\gamma q^2-3\eta H)\frac{\delta\vec{A}^2}{a^2}+\frac{1}{4}q^2\frac{\vec{v}^2}{a^2}\right).
\end{equation}
This shows that $\vec{v}$ is a non dynamical degree of freedom which should be integrated out. Using its EOM, we find that $\vec{v}=0$ and our action reduces to
\begin{equation}\label{vecs}
\mathcal{L}^{(v)}=a^3\left(-2\gamma \frac{\delta\dot{\vec{A}}^2}{a^2}+2 q^2\gamma \frac{\delta\vec{A}^2}{a^2}-3\eta H \frac{\delta\vec{A}^2}{a^2}\right).
\end{equation}
Now imposing the stability conditions given in (\ref{stab}), it is immediately seen that $\gamma<0$ and $\eta>0$. These are the same conditions as we get from the stability of scalar perturbations. It is interesting to see that the mass term in (\ref{vecs}) also has the right sign and is free from tachyonic instabilities.

So far we have  dealt with modes with no dynamics in the background. We have looked at their perturbations and seen that they are completely healthy with regards to stability conditions. Such modes which are trivial at the background and dynamical at perturbation levels are what we expect from isocurvature perturbations. As we saw, the curvature perturbations $\zeta$ are not dynamical as in the case of the original mimetic theory. It should be noted at this point  that $\zeta$  could actually play the role of a dynamical degree of freedom,  much in the same way as in a fully non linear analysis of the original mimetic set up. However, the point is that in cosmology the role played by a dynamical power spectrum (stemming from a quadratic action) is crucial and this is lacking in the framework presented here. What we have done here is to have made the  isocurvature modes  healthy in the context of mimetic theory. Unfortunately the isocurvature modes have no observational contribution to the CMB or LSS and our model still suffers from the absence of propagating curvature perturbations. We shall try to address these problems in the following sections by considering a space-like vector field in an anisotropic universe.

\section{Space-Like Vector Field:\\Bianchi Universe}

Let us now consider the case where a space-like vector $A_a=(0,A_x(t),0,0)$ is present in the background. It is clear that the existence of such a vector field indicates a preferred direction in the background. It is therefore desirable to take a metric of a Bianchi type I form
\begin{equation}\label{biametric}
ds^2=-dt^2+a(t)^2dx^2+b(t)^2(dy^2+dz^2).
\end{equation}
The $tt$, $xx$ and $yy(zz)$ components of the EOM with respect to the metric, equation (\ref{EomFRWg}), are
\begin{equation}\label{biatt}
\begin{split}
tt:\hspace{2mm}2H_aH_b+H_b^2=&-2\lambda-2\gamma\left(\frac{\dot{A_x}}{a}\right)^2\\
&+2\eta\frac{A_x}{a}\frac{\dot{A_x}}{a}-2\eta H_a\left(\frac{A_x}{a}\right)^2,
\end{split}
\end{equation}
\begin{equation}\label{biaxx}
xx: \hspace{2mm}2\dot{H_b}+3H_b^2=-2\gamma\left(\frac{\dot{A_x}}{a}\right)^2-2\eta\frac{A_x}{a}\frac{\dot{A_x}}{a}-4\eta H_b \left(\frac{A_x}{a}\right)^2,
\end{equation}
\begin{equation}\label{biayy}
\begin{split}
yy(zz):&\hspace{2mm}\dot{H_a}+\dot{H_b}+H_a^2+H_b^2+H_a H_b=\\
&+2\gamma\left(\frac{\dot{A_x}}{a}\right)^2+2\eta H_a\left(\frac{A_x}{a}\right)^2-2\eta\frac{\dot{A_x}}{a}\frac{A_x}{a},
\end{split}
\end{equation}
where $H_a\equiv\frac{\dot{a}}{a}$ and $H_b\equiv\frac{\dot{b}}{b} $. In these equations we also use the mimetic constraint, namely $\phi=t$. The EOM for $\phi$, equation(\ref{EomFRWphi}), and the $x$ component of EOM for the vector field, equation (\ref{EomFRWA}), are
\begin{equation}\label{biaphi}
\begin{split}
&\dot{a}a^3 b\lambda +a^4b\dot{\lambda}+2\lambda a^4 \dot{b}-2\eta \dot{a}^2bA_x^2-2\eta a^2 \dot{b}A_x\dot{A_x}-\eta a^2b\dot{A_x}^2\\
&-\eta a^2 b\ddot{A_x}A_x+2\eta\dot{a}a\dot{b}A_x^2+3\eta\dot{a}a b \dot{A_x}A_x+\eta \ddot{a}a b A_x^2=0,
\end{split}
\end{equation}
and
\begin{equation}\label{biaA}
4\gamma\ddot{A_x}-4\gamma(H_a-2H_b)\dot{A_x}-2\eta(H_a+2H_b)A_x=0.
\end{equation}
An inspection shows that equation (\ref{biaphi}) can be written as a total derivative
\begin{equation}
\partial_t\left(ab^2\lambda-\eta \frac{b^2}{a}\dot{A_x}A_x+\eta \frac{\dot{a}b^2}{a}A_x^2\right)=0,
\end{equation}
so that
\begin{equation}
\lambda=-\frac{c/2}{ab^2}+\eta \frac{\dot{A}_x}{a}\frac{A_x}{a}-\eta H \left(\frac{A_x}{a}\right)^2,
\end{equation}
where $c$ is a constant of integration. Substituting $\lambda$ back in equation (\ref{biatt}), the $tt$ component becomes
\begin{equation}\label{Biatt}
2H_aH_b+H_b^2=-2\gamma\left(\frac{\dot{A_x}}{a}\right)^2+\frac{c}{ab^2}.
\end{equation}
The right hand side of this equation is the energy density and that of equations (\ref{biaxx}) and (\ref{biayy}) are minus pressure in the direction of $x$ and $y(z)$. We see that one has a behavior akin to that of matter plus a contribution from the vector field. We shall derive the equation of state for the vector field in what follows and provide some motivations as to its interpretation  as DE.

For a numerical treatment of equations, it is desirable to write the them in terms of dimensionless parameters. By looking at the first term of (\ref{action}) it becomes clear that we may take $\frac{1}{8\pi G}=1$  by re-scaling our coordinates. That is, for having the right mass dimension in the first term of (\ref{action}) our coordinates should have mass dimension of $[x]=\frac{1}{m^2}$. From this we can get other mass dimensions in the action, that is, $[\partial]=m^2$, $[A_a]=1$, $[\phi]=\frac{1}{m^2}$, $[\gamma]=1$ and $[\eta]=m^2$. Now the dimensionless parameters become
\begin{equation}
\tau=H_0 t, \eta=H_0\tilde{\eta}, H_a=H_0 h_a, H_b=H_0 h_b.
\end{equation}
To have an understanding of various limiting behaviors of the model we introduce
\begin{eqnarray}
h\equiv\frac{1}{3}(h_a+2h_b),\\
\sigma\equiv\frac{2}{3}(h_b-h_a),
\end{eqnarray}
where $h$ specifies the average Hubble rate expansion and $\sigma$ indicates deviation from isotropy. For more convenience we also re-scale the vector field
\begin{equation}
X\equiv\frac{A_x}{a}.
\end{equation}
Writing  equations (\ref{biaxx}), (\ref{biayy}), (\ref{biaA}) and (\ref{Biatt}) in terms of these new parameters we find the various components as follows
\begin{equation}\label{eqtt}
3h^2=\frac{3}{4}\sigma^2-2\gamma (X'+(h-\sigma)X)^2+\frac{c/H_0^2}{ab^2},
\end{equation}
\begin{equation}\label{eqxx}
\begin{split}
\hspace{2mm}2h'+&3h^2+\sigma'+3h\sigma+\frac{3}{4}\sigma^2=-2\gamma (X'+(h-\sigma)X)^2\\
&-2\tilde{\eta} (X'+(h-\sigma)X)X-4\tilde{\eta}\left(h+\frac{\sigma}{2}\right)X^2,
\end{split}
\end{equation}
\begin{equation}\label{eqyy}
\begin{split}
\hspace{2mm}2h'&+3h^2-\frac{\sigma'}{2}+\frac{3}{4}\sigma^2-\frac{3}{2}h\sigma=2\gamma (X'+(h-\sigma)X)^2\\
&+2\tilde{\eta}(h-\sigma)X^2-2\tilde{\eta} (X'+(h-\sigma)X)X,
\end{split}
\end{equation}
\begin{equation}\label{veceom}
X''+3hX'+X\left(-\sigma^2+(h'-\sigma')+2 h^2-\frac{3}{2}\frac{\tilde{\eta}}{\gamma} h- h\sigma\right).
\end{equation}
where prime represents derivative with respect to $\tau$. Adding equation (\ref{eqxx}) to 2 times equation (\ref{eqyy}) we  have
\begin{equation}\label{eqii}
\begin{split}
&2h'+3h^2=-\frac{3}{4}\sigma^2+\frac{2}{3}\gamma(X'+(H-\sigma)X)^2\\
&-2\tilde{\eta}(h'+(h-\sigma)X)X-2\tilde{\eta}\sigma X^2
\end{split}
\end{equation}
The right hand side of (\ref{eqtt}) is the energy density and right hand side of(\ref{eqii})is minus pressure. Ignoring the matter term in energy density, the equation of state for the vector field is derived
\begin{equation}
\begin{split}
w=\frac{P}{\rho}=&\frac{1}{\frac{3}{4}\sigma^2-2\gamma (X'+(h-\sigma)X)^2}\left(\frac{3}{4}\sigma^2+2\tilde{\eta} \sigma^2X^2\right.\\
&\left. -\frac{2}{3}\gamma(X'+(h-\sigma)X)^2+2\tilde{\eta} X(h'+X(h-\sigma))\right).
\end{split}
\end{equation}
Ignoring $\sigma$, in the absence of a mass term, the vector field would then represent pure radiation, that is $w=\frac{1}{3}$, which is what we expect for the equation of state. However, we can see that the mass term has the potential to make $w$ negative in a way that it becomes desirable for a description of DE. In this regard a numerical solution for these equations would make the matter more clear.

\begin{figure}
\includegraphics[width=3in]{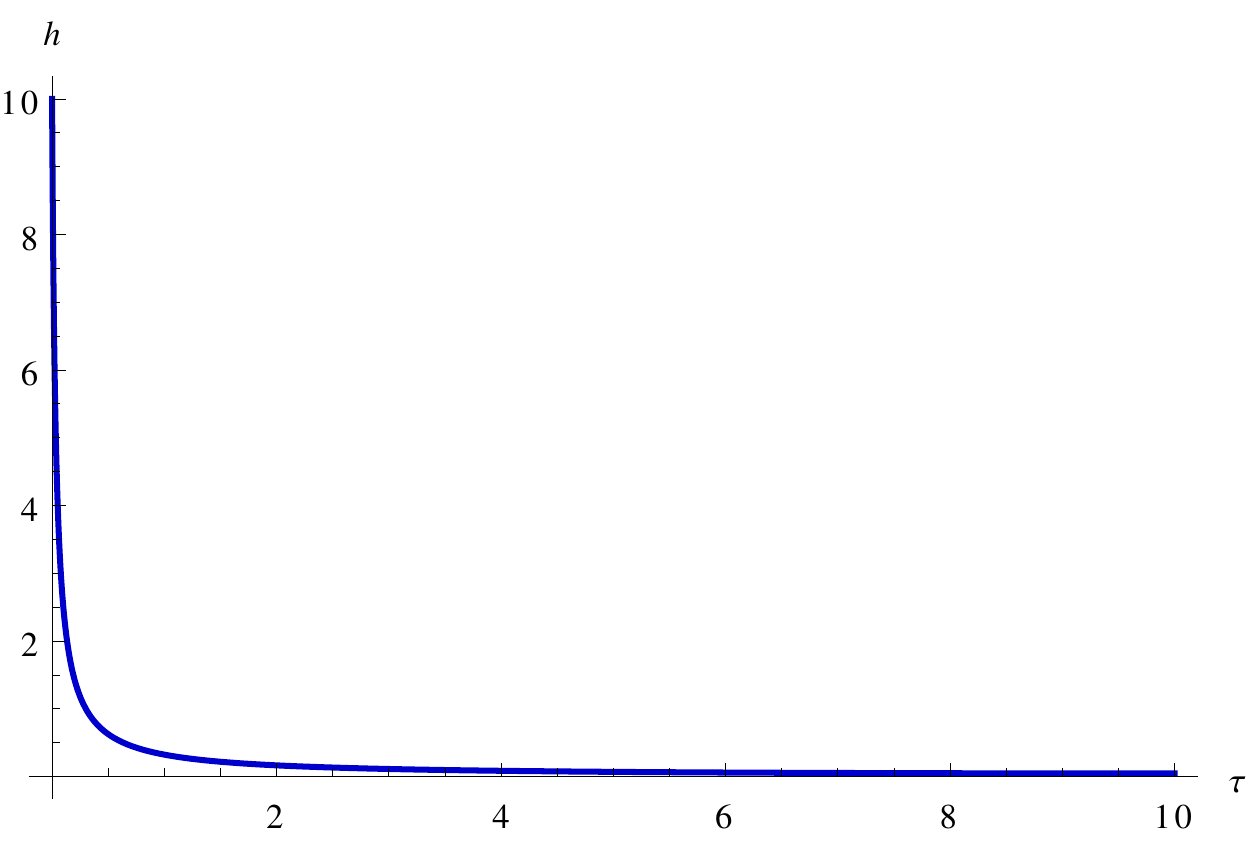}
\caption{\label{BiaSolFig1}\small Numerical solution of the Bianchi background for average Hubble parameter as a function of time with $\gamma=-\frac{1}{4}$ and $\tilde{\eta}=-0.1$ for initial conditions $h(0)=10$, $X(0)=0.1$ and $X'(0)=-0.5$.}
\end{figure}

\begin{figure}
\includegraphics[width=3in]{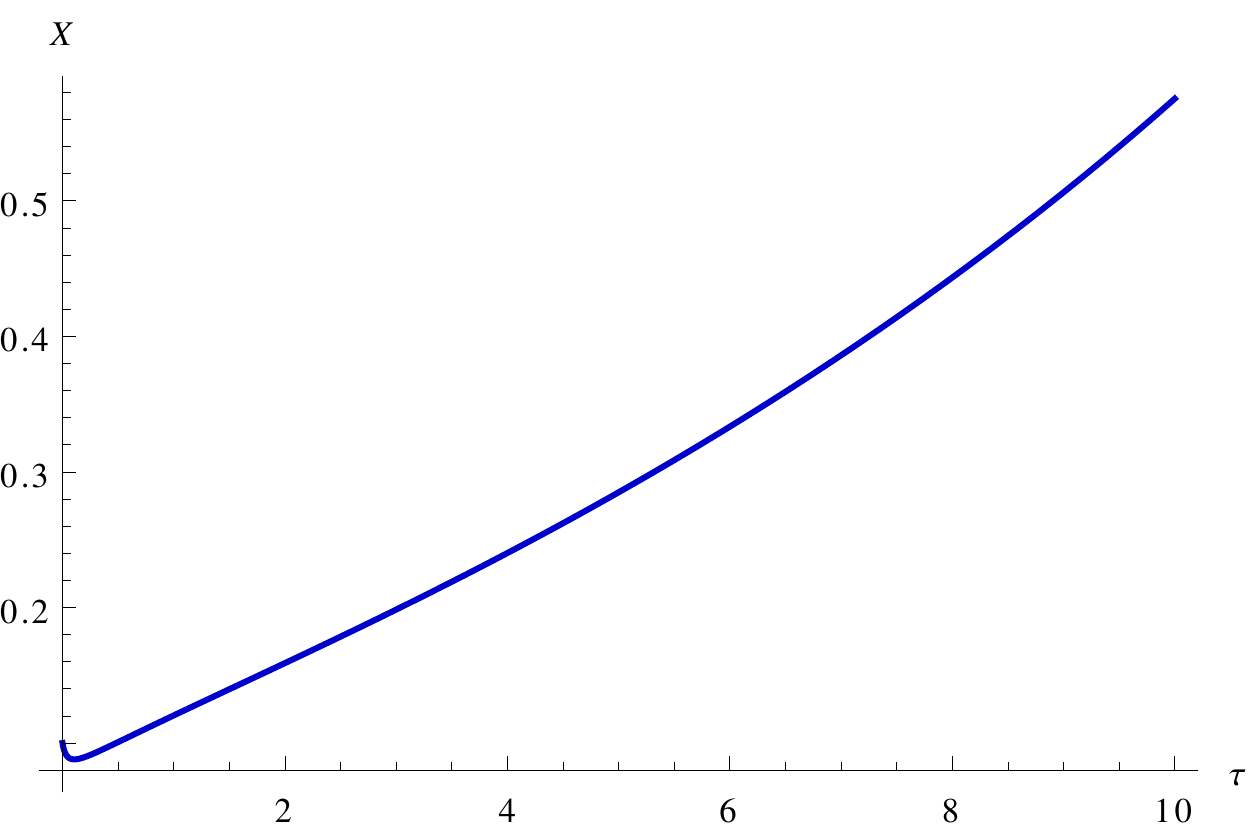}
\caption{\label{BiaSolFig2}\small Numerical solution for scaled component of the vector field $X=A_x/a$ with $\gamma=-\frac{1}{4}$ and $\tilde{eta}=-0.1$ for initial conditions $h(0)=10$, $X(0)=0.1$ and $X'(0)=-0.5$.}
\end{figure}

\begin{figure}
\includegraphics[width=3in]{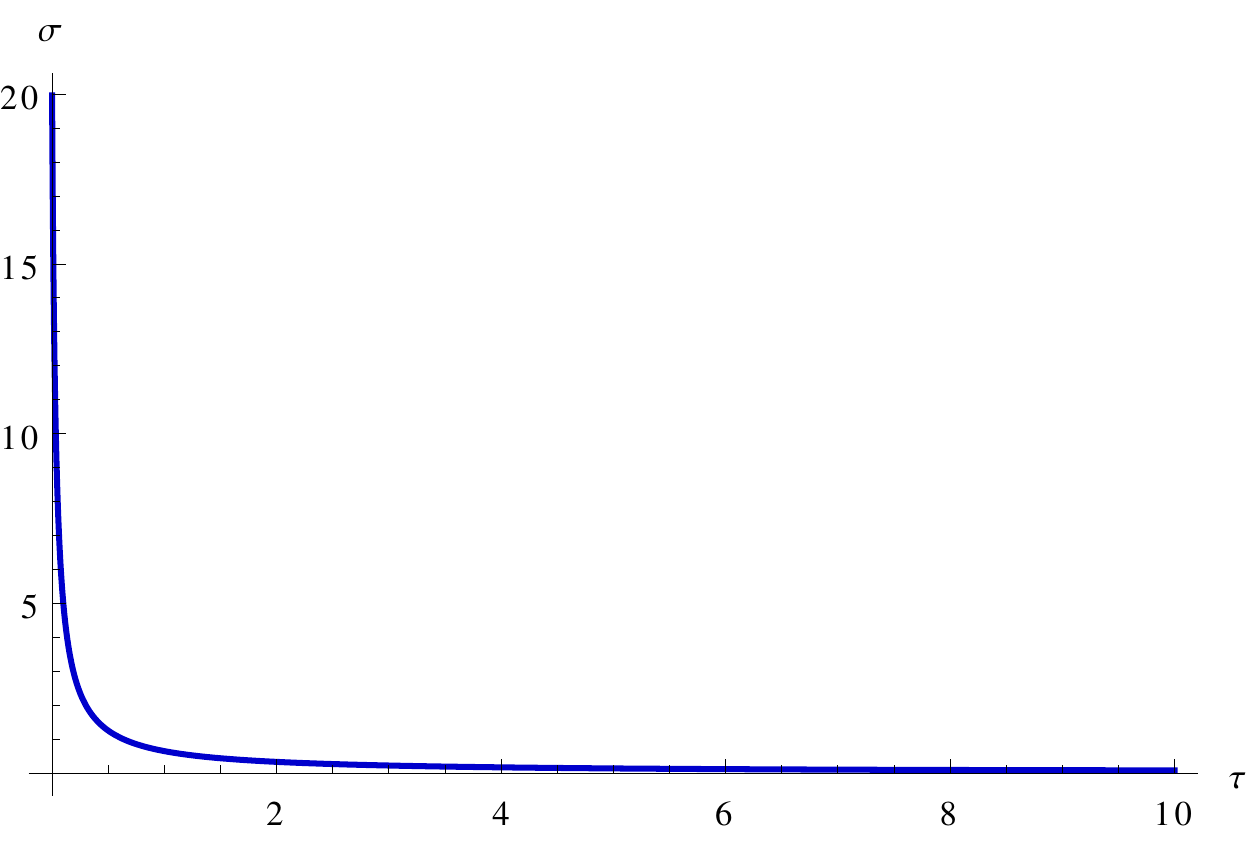}
\caption{\label{BiaSolFig3}Numerical solution for anisotropic parameter $\sigma$ with $\gamma=-\frac{1}{4}$ and $\tilde{\eta}=-0.1$ for initial conditions $h(0)=10$, $X(0)=0.1$ and $X'(0)=-0.5$.}
\end{figure}

\begin{figure}
\includegraphics[width=3in]{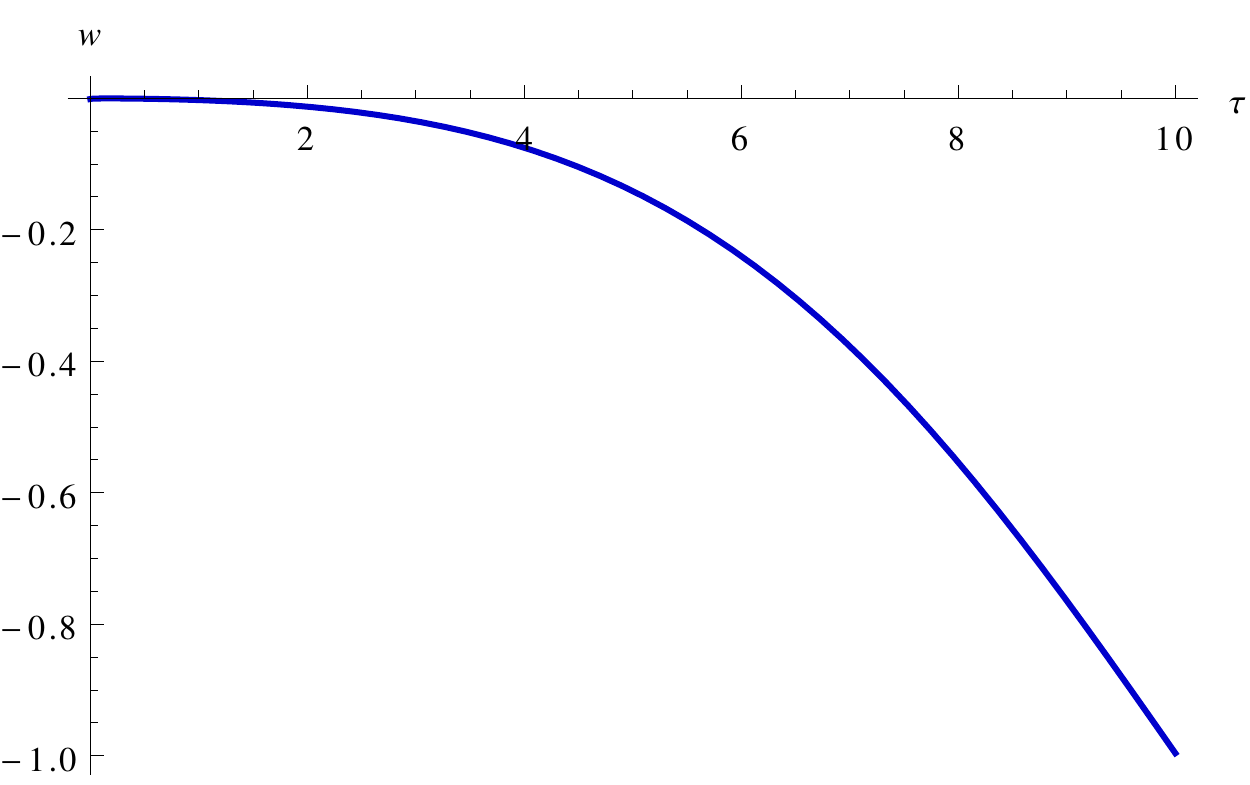}
\caption{\label{BiaSolFig4}Numerical solution for the equation of state parameter $w=p/\rho$ with $\gamma=-\frac{1}{4}$ and $\tilde{\eta}=-0.1$ for initial conditions $h(0)=10$, $X(0)=0.1$and $X'(0)=-0.5$.}
\end{figure}

Ignoring the term $\frac{c/H_0^2}{a^3}$ and solving equation (\ref{eqtt}) in terms of $\sigma$ we find
\begin{equation}
\begin{split}
\sigma=&\frac{1}{8\gamma X^2-3}\Bigl(8\gamma X(X'+hX)\\
&\pm 2\sqrt{3}\sqrt{2\gamma X'(X'+2Xh)+(3-6\gamma X^2)h^2}\Bigr).
\end{split}
\end{equation}
Now, plugging this back to equations (\ref{eqii}) and (\ref{veceom}) results in equations which can be solved numerically. The results for $h$, $X$, $\sigma$  and $w$ are presented in figures \ref{BiaSolFig1}, \ref{BiaSolFig2}, \ref{BiaSolFig3} and \ref{BiaSolFig4}. The plot of $\sigma$ shows that deviation from isotropy decays at lare times.  It is interesting to note that for some reasonable values of $\gamma$ and $\tilde{\eta}$ a desirable DE like equation of state is derived for our vector field.

\section{Stability of Anisotropic Solutions}
The formalism for cosmological perturbations over the background metric given by (\ref{biametric}) was developed in \cite{Gumrukcuoglu:2007bx,Himmetoglu:2008hx,Himmetoglu:2009}. A Bianchi type-I metric has a 2-dimensional isotropy in the $y-z$ plane. We decompose perturbations into scalars and vectors according to the rotational symmetry in the $y-z$ plane. It should be pointed out that no tensor perturbation remains in 2-dimensions. Taking into account the above points we may write the general form of metric perturbations
\begin{equation}
\delta g_{ab}=
\begin{bmatrix}
-2\Phi       & a\partial_x\chi   &b(\partial_iB+B_i)\\
             & -2a^2\Psi         &ab\partial_x(\partial_i\bar{B}+\bar{B}_i)\\
             &                   & b^2(-2\Sigma\delta_{ij}+2\partial_i\partial_jE+\partial_iE_j+\partial_jE_i )\\
\end{bmatrix}.
\end{equation}
The symmetrical components of the metric is not written explicitly so as to make it more readable. Indices $i,j$ stand for $y$ and $z$ directions and $\Phi$, $\chi$, $B$, $\Psi$, $\bar{B}$, $\Sigma$ and $E$ are scalar perturbations and $B_i$, $\bar{B}_i$ and $E_i$ are vector perturbations, which obey transverse conditions
\begin{equation}\label{tr1}
\partial_iB_i=\partial_i\bar{B}_i=\partial_i E_i=0.
\end{equation}
Perturbations of the vector field can be decomposed as
\begin{equation}
\delta A_a=(\delta A_0,\delta A_x,\partial_i\delta A+\delta A_i),
\end{equation}
where $\delta A_0$, $\delta A_x$ and $\delta A$ are scalar perturbations and $\delta A_i$ with $i=y,z$ is the vector perturbation with transverse condition
\begin{equation}\label{tr2}
\partial_i \delta A_i=0.
\end{equation}
It is more convenient to use $SO(2)$ symmetry in the $y-z$ plane, so we can go to a frame where the wave number vector is represented by $\vec{k}=(k_x,k_y,0)$. Unlike the FRW universe, here both longitudinal and transverse modes exist. Also, for our future calculations we define the transverse and longitudinal physical wave numbers as
\begin{equation}
q_x=\frac{k_x}{a},q_y=\frac{k_y}{b}.
\end{equation}
It is now easy to see that in Fourier space, the derivative of modes in the $z$ direction would be zero. Applying this fact to transverse conditions of (\ref{tr1}) and (\ref{tr2}) we see that
\begin{equation}
B_y=\bar{B}_y=E_y=\delta A_y=0.
\end{equation}
Thus in this frame of reference, our scalar and vector perturbations of the metric reduce to
\begin{equation}
\delta g_{ab}^{(s)}=
\begin{bmatrix}
-2\Phi       & a\partial_x\chi   &b\partial_yB                   &0 \\
             & -2a^2\Psi         &ab\partial_x\partial_y\bar{B}         &0 \\
             &                   & b^2(-2\Sigma+2\partial_y^2E)  &0 \\
             &                   &                               &-2b^2\Sigma
\end{bmatrix},
\end{equation}
\begin{equation}
\delta g_{ab}^{(v)}=
\begin{bmatrix}
0      & 0    & 0    &bB_z \\
       & 0    & 0    &ab\partial_x\bar{B}_z \\
       &      & 0    &b^2\partial_y E_z \\
       &      &      &0
\end{bmatrix}.
\end{equation}
Perturbations for the vector field now become
\begin{equation}
\delta A_a^{(s)}=(\delta A_0,\delta A_x,\partial_y \delta A,0),
\end{equation}
\begin{equation}\label{vecvecper}
\delta A_a^{(v)}=(0,0,0,\delta A_z).
\end{equation}

Since there is only the gauge freedom due to diffeomorphism invariance of General Relativity (the $U(1)$ gauge freedom of the vector is broken by the mass term in the action), under general coordinate transformation, $x_a\rightarrow x_a+\xi_a$, the metric transforms as
\begin{equation}
\delta g_{ab}\rightarrow\delta g_{ab}-\partial_c\bar{g}_{ab}\xi^c-\bar{g}_{ac}\partial_b\xi^c-\bar{g}_{bc}\partial_c\xi^c,
\end{equation}
where $\bar{g}$ denotes the background metric. We can parameterize the gauge transformation as $\xi_a=(\xi_0,\partial_x\lambda,\partial_i\Lambda+\xi_i)$, where $\xi_i$ is the 2-dimensional vector degrees of freedom with transverse property, $\partial_i\xi_i=0$, and the others are the 2-dimensional scalar perturbations. Exploiting $SO(2)$ symmetry for the gauge transformation parameters and our preferred reference frame we find
\begin{equation}
\xi_a=(\xi_0,\partial_x \lambda,\partial_y \Lambda,\xi_z).
\end{equation}
Now let us write transformations of perturbation parameters of the metric
\begin{equation}
\begin{split}
&\Phi\rightarrow \Phi -\dot{\xi_0},\\
&\chi\rightarrow \chi+\frac{1}{a}(\frac{1}{a^2}\xi_0-\dot{\lambda}),\\
&B\rightarrow B+b(\frac{1}{b^2}\xi_0-\dot{\Lambda}),\\
&\Psi\rightarrow \Psi+\frac{\dot{a}}{a}\xi_0+\partial_x^2\lambda,\\
&\bar{B}\rightarrow \bar{B}-\frac{b}{a}\Lambda-\frac{a}{b}\lambda,\\
&\Sigma\rightarrow \Sigma +\frac{\dot{b}}{b}\xi_0,\\
&E\rightarrow E-\Lambda,\\
&B_z\rightarrow B_z-b\dot{\xi_z},\\
&\bar{B}_z\rightarrow \bar{B}_z-\frac{b}{a}\xi_z,\\
&E_z \rightarrow E_z-\xi_z.
\end{split}
\end{equation}
In addition, we have the transformation of the scalar field as
\begin{equation}
\delta\phi\rightarrow\delta\phi-\dot{\phi}\xi_0.
\end{equation}
Having these gauge transformations, it is not hard to choose a desirable gauge suitable to the problem at hand. We choose a gauge which is most similar to the comoving gauge in a FRW background. By fixing $\xi_0$ we can go to a gauge where
\begin{equation}
\delta \phi=0.
\end{equation}
As it becomes clear in the following, this choice is preferable for mimetic gravity.
The other three gauge freedoms would be fixed by the choice
\begin{equation}
\bar{B}=E=E_z=0.
\end{equation}
Perturbations now become
\begin{equation}
\delta g_{ab}^{(s)}=
\begin{bmatrix}
-2\Phi       & a\partial_x\chi   &b\partial_yB           &0 \\
             & -2a^2\Psi         &0                      &0 \\
             &                   &-2 b^2\Sigma           &0 \\
             &                   &                       &-2b^2\Sigma
\end{bmatrix},
\end{equation}
and
\begin{equation}\label{metricvecper}
\delta g_{ab}^{(v)}=
\begin{bmatrix}
0      & 0    & 0    &bB_z \\
       & 0    & 0    &ab\partial_x\bar{B}_z \\
       &      & 0    &0\\
       &      &      &0
\end{bmatrix}.
\end{equation}
The procedure is similar to that taken for the FRW case. We decompose our calculations to scalar and vector perturbations and substitute these perturbations in the action which makes it quadratic in terms of perturbations.
There is also a scalar perturbation for the Lagrange multiplier $\delta \lambda$. Keeping this in mind and constructing our quadratic action, the mimetic constraint part of the action is
\begin{equation}
\mathcal{L}\supset ab^2(2\delta \lambda \Phi+\lambda(-4\Sigma\Phi-2\Phi^2-2\Phi\Psi+(\partial_y B)^2+(\partial_x \chi)^2).
\end{equation}
The EOM for $\delta \lambda$ ($\delta\lambda$ only appears in this part of the action, so it is not necessary to be concerned about other terms) leads to $\Phi=0$, i.e. the 00 component of the metric perturbation should be zero. This is what we expect for mimetic gravity perturbations in the comoving gauge. This is one of the reasons for the choice of such a gauge which makes our calculation  much simpler. Knowing this fact and taking $\Phi$ to be zero, the form of the quadratic action in the Fourier space (after using the background equations and some by parts integration for simplification) becomes
\begin{widetext}
\begin{equation}\label{biaL2}
\begin{split}
\mathcal{L}^{(s)}=& ab^2\Biggl(-\dot{\Sigma}^2+q_x^2\Sigma^2-2\dot{\Sigma}\dot{\Psi}+2(-\eta \frac{A_x^2}{a^2}+H_a)\Sigma\dot{\Psi}+2(\eta \frac{A_x^2}{a^2}+H_b)\dot{\Sigma}\Psi+bq_y^2B\dot{\Sigma}+2aq_x^2\dot{\Sigma}\chi+2aq_x^2(-H_a+H_b)\Sigma\chi\\
&+(-6\eta\frac{A_x\dot{A_x}}{a^2}+6\gamma\frac{\dot{A_x^2}}{a^2}+6\eta H_a\frac{A_x^2}{a^2}+2H_aH_b+b^2q_y^2+H_b^2)\Sigma\Psi+8\gamma\frac{\dot{A_x}^2}{a^2}\Sigma\dot{\delta A_x}+4\eta\frac{A_x^2}{a^2}\dot{\Sigma}\delta A_x+4\eta\frac{A_x}{a^2}(H_a+2H_b)\Sigma\delta A_x\\
&-2(\eta\frac{\dot{A_x}A_x}{a^2}+\gamma\frac{\dot{A_x}^2}{a^2}+2\eta H_b\frac{A_x^2}{a^2})\Psi^2+bq_y^2\dot{\Psi}B+bq_y^2(-2\eta \frac{A_x^2}{a^2}+H_a-H_b)\Psi B-2a\eta q_x^2\frac{A_x^2}{a^2}\Psi\chi-4\gamma\frac{\dot{A_x}}{a^2}\Psi\dot{\delta A_x}\\
&+2\eta\frac{A_x}{a^2}\dot{\Psi}\delta A_x-2\eta\frac{A_x}{a^2}(H_a+2H_b)\Psi\delta A_x+\frac{b^2q_x^2q_y^2}{4}B^2-ab\frac{q_x^2q_y^2}{2}B\chi-2bq_y^2(\eta \frac{A_x}{a^2}-2\gamma \frac{\dot{A_x}}{a^2})B\delta A_x-2\eta q_x^2\frac{A_x}{a}\chi\delta A_x\\
&+a^2\frac{q_x^2}{4}(q_y^2+4\eta H_a\frac{A_x^2}{a^2}+8\eta H_b \frac{A_x^2}{a^2})\chi^2+(-2q_x^2\gamma+\eta H_a-2\gamma q_y^2+2\eta H_b)\delta A_0^2-2q_y^2\gamma \dot{\delta A}^2+q_y^2(2q_x^2\gamma-\eta(H_a+2H_b))\delta A^2\\
&+4q_y^2\gamma\delta A_0\dot{\delta A}-2\frac{\gamma}{a^3}\dot{\delta A_x}^2+(2q_y^2\gamma-\eta (H_a+2H_b))\frac{\delta A_x^2}{a^2}\Biggr).
\end{split}
\end{equation}
\end{widetext}
Looking at the above quadratic action we can see that $\chi$, $B$ and $\delta A_0$ do not appear with time derivatives. This means that they are non dynamical degrees of freedom and should be integrated out.

Varing the above action with respect to $\delta A_0$ results in
\begin{equation}
\delta A_0=\frac{\gamma q_y^2\delta\dot{A}}{(q_x^2+q_y^2)\gamma-\frac{\eta}{2}(H_a+2H_b)}.
\end{equation}
The EOM for $\chi$ is
\begin{equation}\label{eqchi}
\begin{split}
a\chi(q_y^2&+4\eta\frac{A_x^2}{a^2}(H_a+2H_b))-bq_y^2B\\
&=4\eta\frac{A_x}{a}\frac{\delta A_x}{a}+4\eta\frac{A_x^2}{a^2}\Psi-4\dot{\Sigma}
+4(H_a-H_b)\Sigma,
\end{split}
\end{equation}
and that of for $B$ becomes
\begin{equation}\label{eqB}
\begin{split}
aq_x^2\chi-bq_x^2B=&4\frac{2\gamma\dot{A_x}-\eta A_x}{a}\frac{\delta A_x}{a}+2(\dot{\Sigma}+\dot{\Psi})\\
&+2(H_a-H_b-2\eta\frac{A_x^2}{a^2})\Psi.
\end{split}
\end{equation}
We can solve equations (\ref{eqchi}) and (\ref{eqB}) to find a solution for $B$ and $\chi$ in terms of  other dynamical modes
\begin{equation}
\begin{split}
a\chi=&f_0\biggl(4(\eta\frac{A_x^2}{a^2}-q_y^2\frac{2\gamma\dot{A_x}/a-\eta A_x/a}{q_x^2})\frac{\delta A_x}{a}-2(2+\frac{q_y^2}{q_x^2})\dot{\Sigma}\\
&+4(H_a-H_b)\Sigma+(4\eta\frac{A_x^2}{a^2}-2\frac{q_y^2}{q_x^2}(H_a-H_b-2\eta \frac{A_x^2}{a^2}))\Psi\\
&-2\frac{q_y^2}{q_x^2}\dot{\Psi}\biggr),
\end{split}
\end{equation}
\begin{equation}
\begin{split}
bB=&4(\eta f_0\frac{A_x^2}{a^2}-\frac{1+q_y^2f_0}{q_x^2}(2\gamma\frac{\dot{A}_x}{a}-\eta\frac{A_x}{a}))\frac{\delta A_x}{a}\\
&+(4\eta f_0 \frac{A_x^2}{a^2}-2\frac{1+q_y^2f_0}{q_x^2}(H_a-H_b-2\eta \frac{A_x^2}{a^2}))\Psi\\
&-(4f_0+2\frac{1+q_y^2 f_0}{q_x^2})\dot{\Sigma}+4f_0(H_a-H_b)\Sigma\\
&-2\frac{1+q_y^2f_0}{q_x^2}\dot{\Psi},
\end{split}
\end{equation}
where we define  $f_0$ as
\begin{equation}
f_0\equiv \frac{1}{4\eta\frac{A_x^2}{a^2}(H_a+2H_b)}.
\end{equation}
Now if we insert these solutions back to equation (\ref{biaL2}), we will get a quadratic action in terms of dynamical modes. The resulting equation is lengthy and complicated. However, since we are interested in the ghost instability, we separate the kinetic terms of the quadratic action with the result
\begin{equation}\label{biakin}
\begin{split}
\mathcal{L}^{(s)}\supset& ab^2\biggl(-\frac{f_0}{q_x^2}(q_y^4+4q_x^2q_y^2+4q_x^4+\frac{1}{f_0}(q_x^2+q_y^2))\dot{\Sigma}^2\\
&-\frac{2f_0}{q_x^2}(q_y^4+2q_x^2q_y^2+\frac{1}{f_0}(q_x^2+q_y^2))\dot{\Sigma}\dot{\Psi}\\
&-\frac{f_0}{q_x^2}(q_y^4+\frac{1}{f_0}q_y^2)\dot{\Psi}^2-2\gamma\frac{\delta\dot{A}_x^2}{a^2}\\
&-2q_y^2\frac{2\gamma^2 q_x^2-\eta\gamma(H_a+2H_b)}{2\gamma(q_x^2+q_y^2)-\eta(H_a+2H_b)}\delta\dot{A}^2\biggr).
\end{split}
\end{equation}
What is really malignant is the existence of the Ghost instability at UV. As pointed out in \cite{Gumrukcuoglu:2016jbh, Emami:2016ldl}, the existence of ghost instability in the IR regime indicates a Jeans-like instability that only affects the long wavelength modes. So the existence of low momentum ghost instability does not indicate a pathology in a theory. With this in mind, we can construct our Kinetic matrix in the UV limit as
\begin{equation}
\mathcal{L}^{(s)}_{UV}\supset ab^2
\begin{bmatrix}
\dot{\Sigma}    & \dot{\Psi }  &\dot{\delta A}      &\dot{\delta A_x}
\end{bmatrix}
\bold{K_{UV}}
\begin{bmatrix}
&\dot{\Sigma}\\
&\dot{\Psi}\\
&\dot{\delta A}\\
&\dot{\delta A_x}
\end{bmatrix}.
\end{equation}
\begin{equation}
\bold{K_{UV}}=\begin{bmatrix}
-\frac{f_0}{q_x^2}(q_y^2+2q_x^2)^2   & -\frac{f_0}{q_x^2}q_y^2(q_y^2+2q_x^2)  &     &   \\
-\frac{f_0}{q_x^2}q_y^2(q_y^2+2q_x^2)    &-\frac{f_0}{q_x^2}q_y^4 &    &  \\
&     &     &-2\gamma\frac{q_x^2q_y^2}{q_x^2+q_y^2}  &  \\
&     &     &      &-2\frac{\gamma}{a^2}
\end{bmatrix}.
\end{equation}
The two diagonal terms in the kinetic matrix are seen to satisfy the condition (\ref{stab}) for avoiding ghosts  for $\gamma<0$. As for the other two, we  diagonalize the corresponding matrix for which the  eigenvalues are
\begin{equation}
 0, -2 \frac{f_0}{q_x^2}(q_y^4+2q_x^2q_y^2+2q_x^4).
\end{equation}
We see that  the sign of nonzero eigenvalue is fixed by $f_0$ and therefore it would be free from ghost at UV for $\eta<0$. It is interesting to note that in the Bianchi case  we still have one non propagating curvature perturbation. However, we have succeeded to turn the other curvature perturbation on,  which at least, is free from ghost at UV.
The other parts that relate to gradient instability are given in the appendix. It is interesting to note that in the UV regime only one term
\begin{equation}
\mathcal{L}_{UV}^{(s)}\supset 2q_x^2q_y^2\gamma \delta A^2,
\end{equation}
is dominant which is free from gradient instability for $\gamma<0$.

To complete our analyses of the model we investigate  vector perturbations. Substituting  perturbations  of the vector part of the metric and vector field, equations (\ref{metricvecper}) and (\ref{vecvecper}) into the action, its quadratic form in Fourier space, after some simplifications by background equations and integration by parts becomes
\begin{equation}\label{actionvec}
\begin{split}
&\mathcal{L}^{(v)}=ab^2\Biggl(\frac{1}{4}(q_x^2+q_y^2)B_z^2-\frac{a}{2}q_x^2B_z\dot{\bar{B}}_z-\frac{a}{2}q_x^2(H_a-H_b)B_z\bar{B}_z\\
&+\frac{a^2}{4}q_x^2\dot{\bar{B}}_z^2-2\gamma \frac{\delta\dot{A}_z^2}{b^2}-\frac{1}{4}q_x^2(2\eta (H_a +2H_b)\frac{A_x^2}{a^2}+4\gamma\frac{\dot{A}_x^2}{a^2}+q_y^2\\
&-(H_a-H_b)^2)\bar{B}_z^2+(2 \gamma(q_x^2+q_y^2)-\eta(H_a+2 H_b))\frac{\delta A_z^2}{b^2}\Biggr).
\end{split}
\end{equation}
This shows that the mode $B_z$ is non dynamical and should be integrated out. The EOM for $B_z$ is
\begin{equation}
(q_x^2+q_y^2)B_z-aq_x^2(H_a+H_b)\bar{B}_z-aq_x^2\dot{\bar{B}}_z=0,
\end{equation}
from which  $B_z$ in terms of $\bar{B}_z$ is given by
\begin{equation}
B_z=\frac{a q_x^2}{q_x^2+q_y^2}((H_a+H_b)\bar{B}_z+\dot{\bar{B}}_z).
\end{equation}
Inserting the solution back to (\ref{actionvec}), we get the quadratic action
\begin{equation}
\begin{split}
\mathcal{L}^{(v)}=&ab^2\biggr(\frac{a^2 q_x^2q_y^2}{4(q_x^2+q_y^2)}\dot{\bar{B}}_z^2-g_1 \bar{B}_z^2-2\gamma\frac{\delta\dot{A}_z^2}{b^2}\\
&+(2\gamma(q_x^2+q_y^2)-\eta (H_a+2H_b))\frac{\delta A_z^2}{b^2}\biggr),
\end{split}
\end{equation}
where we have defined $g_1$ as
\begin{equation}
\begin{split}
g_1\equiv \frac{a^2}{4}\frac{q_x^2q_y^2}{q_x^2+q_y^2}&\biggl( 2\eta\frac{A_x^2}{a^2}(H_a+2H_b)+4\gamma\frac{\dot{A}_x^2}{a^2}+q_x^2+q_y^2\\
&+(H_a-H_b)^2\frac{q_x^2-q_y^2}{q_x^2+q_y^2}\biggr).
\end{split}
\end{equation}
This indicates that the vector perturbations for $\gamma<0$ are free from ghost in general and avoid gradient instability, at least in the high momenta regime.

\section{Conclusions}
We have studied a model for mimetic gravity where a vector field is added to the Lagrangian which is coupled to the mimetic field through the mass term of the vector field, that is $\Box\phi$ plays the role of a mass term for our Proca-like vector field. We studied the model in two cases,  a  time-like vector field which corresponds to FRW universe and a space-like vector field which corresponds to a Bianchi universe. We also studied the stability of the resulting FRW universe and showed that although a propagating curvature perturbation is absent, a healthy isocurvature mode can be turned on. We also looked for healthy curvature perturbations and  investigated the case of a  Bianchi universe. Knowing that serious pathology of a  theory exhibits itself only at short wavelength modes, we checked the stability at the UV limit. We showed that vector perturbations are free from ghosts in general and free from gradient instabilities at high momenta, whereas scalar perturbations are free from ghosts and gradient instabilities at UV.

It would be interesting to perform numerical analysis to check if such stabilities remain at all order of momenta. Also desirable is analyzing other kinds of vector scalar mimetic Lagrangians with different kind of interactions which have more phenomenological basis. We will address these questions in future works.
\begin{acknowledgments}
The authors would like to thank Mohammad Ali Gorji for helpful discussions through out the work and Ghadir Jaffari for his help with xAct package \cite{Brizuela:2008ra,Nutma:2013zea} and  helpful comments. We also thank S. Shahdi for his careful reading of the manuscript and useful comments.
\end{acknowledgments}

\appendix
\section{Further Details}
The quadratic action of scalar perturbations of FRW case, in real space, before equating the laps function to zero
\begin{widetext}
\begin{small}
\begin{eqnarray}
&&\mathcal{L}^{(s)}=- \dot{a}\bigl(3 \eta (\frac{\partial \delta A}{\partial z})^2 + 3 \frac{\partial \mathcal{N}}{\partial z} \frac{\partial \chi}{\partial z} \nonumber + 2 \frac{\partial \zeta}{\partial z} \frac{\partial \
\chi}{\partial z} + 2 \mathcal{N} \frac{\partial^{2}\chi}{\partial \
z^{2}} - 2 \zeta \frac{\partial^{2}\chi}{\partial z^{2}} \nonumber + 3 \eta (\frac{\partial \delta A}{\partial y})^2 + 3\frac{\partial \mathcal{N}}{\partial y} \frac{\partial \chi}{\partial y} - 2 \frac{\partial \zeta}{\partial y} \frac{\partial \chi}{\partial y} \nonumber + 2 \mathcal{N} \frac{\partial^{2}\chi}{\partial y^{2}} - 2 \zeta \frac{\partial^{2}\chi}{\partial y^{2}} \\
&&+ 3 \eta (\frac{\partial \delta A_x}{\partial x})^2 + 3 \frac{\partial \mathcal{N}}{\partial x} \frac{\partial \chi}{\partial x} \nonumber - 2 \frac{\partial \zeta}{\partial x} \frac{\partial \chi}{\partial x} + 2 \mathcal{N} \frac{\partial^{2}\chi}{\partial
x^{2}} - 2 \zeta \frac{\partial^{2}\chi}{\partial x^{2}}\bigr) + \frac{1}{a}((\frac{\partial^{2}\chi}{\partial z^{2}})^2 + (\frac{\partial^{2}\chi}{\partial y\partial z})^2 + \frac{\partial\chi}{\partial y}\frac{\partial^{3}\chi}{\partial y\partial z^{2}} + (\frac{\partial^{2}\chi}{\partial y^{2}})^2 + \frac{\partial \chi}{\partial y} \frac{\partial^{3}\chi}{\partial y^{3}}\\
&& + (\frac{\partial^{2}\chi}{\partial x\partial z})^2 + \frac{\partial \chi}{\partial x} \frac{\partial^{3}\chi}{\partial x\partial z^{2}} + (\frac{\partial^{2}\chi}{\partial x\partial y})^2 + \frac{\partial
\chi}{\partial x} \frac{\partial^{3}\chi}{\partial x\partial y^{2}} +
\frac{\partial^{2}\chi}{\partial y^{2}} \frac{\partial^{2}\chi}{\partial x^{2}} + (\frac{\partial^{2}\chi}{\partial x^{2}})^2 + \frac{\partial^{2}\chi}{\partial z^{2}} (\frac{\partial^{2}\chi}{\partial y^{2}} +\frac{\partial^{2}\chi}{\partial x^{2}}) + \frac{\partial \chi}{\partial z} (\frac{\partial^{3}\chi}{\partial z^{3}} + \frac{\partial^{3}\chi}{\partial y^{2}\partial z}\\
&& + \frac{\partial^{3}\chi}{\partial x^{2}\partial z}) + \frac{\partial \chi}{\partial y} \frac{\partial^{3}\chi}{\partial x^{2}\partial y} + \frac{\partial \chi}{\partial x} \frac{\partial^{3}\chi}{\partial x^{3}}) + \tfrac{3}{2} (\mathit{a})^2 \Bigl(2 \eta (\delta A_0)^2 \
\dot{a} + 2\mathcal{N}^2 \ddot{a} + 3 \zeta \bigl(3 \zeta \
\ddot{a} - 2 \dot{a} (\dot{\mathcal{N}}- 4 \dot{\zeta})\bigr)+ \mathcal{N} \bigl(-6 \zeta \ddot{a}\nonumber + 4 \dot{a} (\dot{\mathcal{N}} - 2 \dot{\zeta}\bigr)\Bigr) +\\
&& \tfrac{1}{2} \mathit{a} \Bigl(6 (\mathcal{N})^2 \dot{a}^2 \nonumber  + 27 (\zeta)^2 \dot{a}^2 - 2 \
\bigl(2 \gamma (\frac{\partial \delta A0}{\partial z})^2 + \
\frac{\partial \mathcal{N}}{\partial z} \frac{\partial \
\zeta}{\partial z} \nonumber + (\frac{\partial \zeta}{\partial z})^2 + 2 \gamma (\frac{\partial \
\delta A0}{\partial y})^2 + \frac{\partial \mathcal{N}}{\partial y} \
\frac{\partial \zeta}{\partial y} + (\frac{\partial \zeta}{\partial \
y})^2 \nonumber\\
&& + 2 \gamma (\frac{\partial \delta A_0}{\partial x})^2 + \
\frac{\partial \mathcal{N}}{\partial x} \frac{\partial \
\zeta}{\partial x}+ (\frac{\partial \zeta}{\partial x})^2 \nonumber + \frac{\partial^{2}\chi}{\partial z^{2}} \dot{\mathcal{N}} + \frac{\partial^{2}\chi}{\partial y^{2}} \
\dot{\mathcal{N}}\nonumber + \frac{\partial^{2}\chi}{\partial x^{2}} \dot{\mathcal{N}} - 2 \frac{\partial^{2}\chi}{\partial z^{2}} \
\dot{\zeta} - 2 \
\frac{\partial^{2}\chi}{\partial y^{2}} \dot{\zeta}\nonumber- 2 \frac{\partial^{2}\chi}{\partial x^{2}}\dot{\zeta} - 4 \gamma \frac{\partial \delta A_0}{\partial z} \
\frac{\partial\dot{\delta A}}{\partial z} \nonumber \\
&& + 2 \gamma (\frac{\partial\dot{\delta A}}{\partial z})^2 - \
4 \frac{\partial \chi}{\partial z} \frac{\partial\dot{\zeta}}{\partial z} -  \frac{\partial \zeta}{\partial z} \
\frac{\partial\dot{\chi}}{\partial z} \nonumber - 4 \gamma \frac{\partial \delta A0}{\partial y} \
\frac{\partial\dot{\delta A}}{\partial y} + 2 \gamma \
(\frac{\partial\dot{\delta A}}{\partial y})^2 \nonumber - 4 \frac{\partial \chi}{\partial y} \
\frac{\partial\dot{\zeta}}{\partial y} -  \frac{\partial \
\zeta}{\partial y} \frac{\partial\dot{\chi}}{\partial y} \nonumber - 4 \gamma \frac{\partial \delta A0}{\partial x} \
\frac{\partial\dot{\delta A}}{\partial x}\\
&& + 2 \gamma (\frac{\partial\dot{\delta A}}{\partial x})^2 \nonumber - 4 \frac{\partial \chi}{\partial x} \
\frac{\partial\dot{\zeta}}{\partial x} -  \frac{\partial \
\zeta}{\partial x} \frac{\partial\dot{\chi}}{\partial \
x}\bigr) \nonumber - 2 \zeta (\frac{\partial^{2}\mathcal{N}}{\partial z^{2}} + 2 \frac{\partial^{2}\zeta}{\partial z^{2}} + \frac{\partial^{2}\mathcal{N}}{\partial y^{2}} + 2 \frac{\partial^{2}\zeta}{\partial y^{2}} \nonumber + \frac{\partial^{2}\mathcal{N}}{\partial x^{2}} + 2 \
\frac{\partial^{2}\zeta}{\partial x^{2}} -  \
\frac{\partial^{2}\dot{\chi}}{\partial z^{2}} -  \
\frac{\partial^{2}\dot{\chi}}{\partial y^{2}} \nonumber -  \frac{\partial^{2}\dot{\chi}}{\partial x^{2}})\\
&& - 2 \mathcal{N} \bigl(9 \zeta \dot{a}^2 \
+ 2 \frac{\partial^{2}\zeta}{\partial z^{2}} + 2 \frac{\partial^{2}\zeta}{\partial y^{2}} \nonumber  + 2 \frac{\partial^{2}\zeta}{\partial x^{2}} + \
\frac{\partial^{2}\dot{\chi}}{\partial z^{2}} + \frac{\partial^{2}\dot{\chi}}{\partial y^{2}} + \frac{\partial^{2}\dot{\chi}}{\partial x^{2}}\bigr)\Bigr) \nonumber -  (\mathit{a})^3 \bigl((\mathcal{N})^2 \lambda + 3 \dot{\mathcal{N}} \dot{\zeta}- 6 (\dot{\zeta})^2 - 9 \zeta \ddot{\zeta} \nonumber \\
&& + \mathcal{N} (-2 \delta \lambda - 6 \zeta \lambda + 3\ddot{\zeta})\nonumber \bigr).
\end{eqnarray}
\end{small}
\end{widetext}
The quadratic action for scalar perturbations in Fourier space relevant to gradient instability of the Bianchi case
\begin{widetext}
\begin{equation}
\begin{split}
&\mathcal{L}^{(s)}_{Gradinet}=ab^2\Biggl(\frac{\Sigma^2}{(H_a+2H_b)^2}\biggl(q_x^2\Bigl(3\frac{\dot{A}_x/a}{A_x/a}(H_b-H_a)+\frac{1}{2\eta}\frac{\dot{A}_x/a}{A_x^2/a^3}(H_a^2+H_aH_b-2H_b^2)+H_a^2-8H_aH_b-2H_b^2\\
&-3\frac{\gamma}{\eta}\frac{\dot{A}_x^2/a^2}{A_x^2/a^2}-\frac{2}{\eta}\frac{H_a^3}{A_x^2/a^2}-\frac{9\gamma}{\eta}\frac{\dot{A}_x^2/a^2}{A_x^2/a^2}H_b^2+\frac{3}{2\eta}\frac{H_aH_b^2}{A_x^2/a^2}+\frac{1}{2\eta}\frac{H_b^3}{A_x^2/a^2}\Bigr)+q_y^2\Bigl(\frac{3}{2}\frac{\dot{A}_x/a}{A_x/a}(H_b-H_a)+\frac{1}{4\eta}\frac{\dot{A}_x/a}{A_x^2/a^3}(H_a^2+H_aH_b-2H_b^2)\\
&-6H_aH_b-3H_b^2-\frac{3\gamma}{2\eta}\frac{\dot{A}_x^2/a^2}{A_x^2/a^2}H_a-\frac{3}{2\eta}\frac{H_a^3}{A_x^2/a^2}+\frac{9}{4\eta}\frac{H_aH_b^2}{A_x^2/a^2}-\frac{3}{4\eta}\frac{H_b^3}{A_x^2/a^2}-\frac{\gamma}{2\eta}H_b\Bigr) \biggr)+\frac{\dot{\Sigma}\Psi}{2\eta A_x^2/a^2(H_a+2H_b)}\biggl(4q_y^2(H_b-H_a)\\
&\frac{q_y^4}{q_x^2}(H_b-H_a)+4\eta q_x^2\frac{A_x^2}{a^2}+4\eta \frac{A_x^2}{a^2}\frac{q_y^4}{q_x^2}+6\eta\frac{A_x^2}{a^2}q_y^2\biggr)+\frac{\Sigma\Psi}{2\eta A_x^3/a^3(H_a+2H_b)^2}\Bigl(6q_y^2\frac{A_x^2}{a^2}\frac{\dot{A}_x}{a}(H_b-H_a)+4q_y^2\frac{\dot{A}_x}{a}(H_a^2+H_aH_b-2H_b^2)\\
&q_y^2\frac{A_x}{a}(-2(3\gamma\frac{\dot{A}_x^2}{a^2}H_a+2H_a^3)-18\gamma\frac{\dot{A}_x^2}{a^2}H_b+3H_aH_b^2+H_b^3)+2\eta \frac{A_x^3}{a^3}(2q_x^2H_a^2+10q_y^2H_aH_b+2q_x^2H_aH_b-2q_y^2H_b^2+q_y^2H_a^2-4q_x^2H_b^2 )\Bigr)\\
&+\frac{\dot{\Sigma}\delta A_x}{\eta A_x^2/a^2(H_a+2H_b)}\Bigl(\eta \frac{A_x}{a}(\frac{q_y^4}{q_x^2}+3q_y^2+2q_x^2)-2\gamma \frac{q_y^2}{q_x^2}(q_y^2+2q_x^2) \Bigr)+\frac{2\Sigma \delta A_x(H_b-H_a)}{\eta A_x^2/a^2(H_a+2H_b)}\Bigl(\eta\frac{A_x}{a}(q_x^2+q_y^2)-2q_y^2\gamma\frac{\dot{A}_x}{a}\Bigr)\\
&\frac{\dot{\Psi}\delta A_x}{\eta A_x^2/a^2(H_a+2H_b)}(\frac{q_y^2}{q_x^2}\eta\frac{A_x}{a}(q_x^2+q_y^2)-2\gamma\frac{\dot{A}_x}{a}\frac{q_y^4}{q_x^2})+\frac{\Psi\delta A_x}{\eta A_x^2/a^2(H_a+2H_b)}\Bigl(-\frac{q_y^2}{q_x^2}\eta\frac{A_x}{a}(q_x^2+q_y^2)(H_b-H_a)+2\frac{q_y^4}{q_x^2}\gamma\frac{\dot{A}_x}{a}(H_b-H_a)\\
&+4\frac{q_y^2}{q_x^2}\gamma\eta\frac{A_x^2}{a^2}\frac{\dot{A}_x}{a}(q_x^2+q_y^2)-2\eta^2\frac{A_x^3}{a^3}(q_x^2+2q_y^2+\frac{q_y^4}{q_x^2})\Bigr)+\frac{\delta A_x^2}{\eta A_x^2/a^2(H_a+2H_b)}\Bigl(-\eta\frac{A_x^2}{a^2}(\eta\frac{q_y^4}{q_x^2}+\eta q_x^2+2\gamma H_a q_y^2-4\gamma H_bq_y^2+\eta q_y^2)\\
&-4\gamma^2\frac{q_y^4}{q_x^2}\frac{\dot{A}_x^2}{a^2}+4\frac{q_y^2}{q_x^2}\gamma\eta\frac{A_x}{a}(q_y^2+q_x^2) \Bigr)+q_y^2\Bigl(2q_x^2\gamma-\eta (H_a+2H_b)\Bigr)\delta A^2+\frac{\Psi^2}{8q_x^2\eta A_x^3/a^3(H_a+2H_b)}\Bigl( 6\frac{q_y^4}{q_x^2}\frac{A_x^2}{a^2}\frac{\dot{A}_x}{a}(H_a-H_b)\\
&4\frac{q_y^4}{q_x^2}\frac{\dot{A}_x}{a}(-H_b^2-H_aH_b+2H_b^2)+\frac{q_y^4}{q_x^2}\frac{A_x}{a}(6\gamma\frac{\dot{A}_x^2}{a^2}H_a+8H_a^3+18\gamma \frac{\dot{A}_x^2}{a^2}H_b-15H_aH_b^2+7H_b^3)+12\eta^2\frac{A_x^4}{a^4}\frac{\dot{A}_x}{a}(\frac{q_y^4}{q_x^2}+q_y^2)\\
&-8\eta^2\frac{A_x^5}{a^5}((H_a+2H_b)q_x^2+3H_b\frac{q_y^4}{q_x^2}+q_y^2H_a+5q_y^2)+2\eta\frac{A_x^3}{a^3}(-2q_y^2H_aH_b+6\frac{q_y^4}{q_x^2}H_aH_b +H_b^2\frac{q_y^4}{q_x^2}+q_y^2(-13H_b^2-4H_a^2\frac{q_y^2}{q_x^2}\\
&+2\gamma\frac{\dot{A}_x^2}{a^2}\frac{q_y^2}{q_x^2})+2\gamma\frac{\dot{A}_x^2}{a^2}q_y^2)\Bigr) \Biggr).
\end{split}
\end{equation}
\end{widetext}

\bibliography{apssamp5}

\end{document}